\documentclass[twocolumn,twoside,slac_two]{revtex4}
\usepackage{graphicx}
\usepackage{fancyhdr}
\usepackage{amsmath,amssymb}
\newcommand{\CA}{{\cal A}}
\newcommand{\nn}{\nonumber}
\begin{document}

\title{Theory Overview}

%

\author{Benjam\'\i{}n Grinstein}
\affiliation{Physics Department, University of California, San Diego; La Jolla, CA
 92093-0319, USA
}%

\begin{abstract}
We first review some aspects of the determination of the sides and
angles of the unitarity triangle. We pay particular attention to
theory shortcomings, and present many alternative proposals for the
determination of $|V_{ub}|$ (which at present is problematic). We then
turn our attention to the more general question: What have we learned
so far about flavor physics and where do we go from here? We argue that
the aim of Flavor Physics should be to establish or rule out Minimal
Flavor Violating interactions up to a scale of 10~TeV.  
\end{abstract}

\maketitle

\thispagestyle{fancy}


\section{Introduction and UT Theory}
\label{sec:sides}
In this first section of this talk I will skim over the determination
of sides and angles of the unitarity triangle (UT). I do not pretend
to make a complete review or even an overview. I picked topics on the
basis of where I thought we should be weary overly cautions of
theorists ``predictions.''  In subsequent sections I attempt to get
some perspective on the field, and will ask and try to answer the
questions of 
what we have learned in FP and CP physics and where should we go from
here.

\subsection{$|V_{td}/V_{ts}|$}
\label{vtdvts}
The magnitudes of $V_{td}$ and $V_{ts}$ are determined from
measurements of neutral $B_d$ and $B_s$ oscillations,
respectively. The big news last year was the precise measurement of
the $B_s$ mixing rate at Tevatron
experiments\cite{Abulencia:2006mq,Abulencia:2006ze}. While $|V_{ts}|$
does not provide direct information on the apex of the unitarity
triangle, the ratio $|V_{td}/V_{ts}|$ does. The interest in the ratio
stems from the cancellation of hadronic uncertainties:
\begin{equation}
\frac{|V_{td}|}{|V_{ts}|}=\xi\sqrt\frac{\Delta m_s\, m_{B_s}}{\Delta m_d\,
  m_{B_d}}, \quad\text{where}\quad
\xi^2\equiv\frac{B_{B_s}f_{B_s}^2}{B_{B_d}f_{B_d}^2}.
\end{equation}
The hadronic parameter $\xi$ would be unity in the flavor-$SU(3)$
symmetry limit. Lattice QCD gives\cite{Aoki:2003xb}
$\xi=1.21{\textstyle{+0.047\atop-0.035}}$, and  combining with the
experimental result  
\[\frac{|V_{td}|}{|V_{ts}|}=0.2060\pm
0.0007\text{(exp)}{\textstyle{+0.0081\atop-0.0060}}\text{(theory)}
\]
The error, approximately 3\%, is dominated by theory, which comes
solely from the error in $\xi$. There aren't many examples of quantities
that the lattice has post-dicted (let alone predicted) with this sort
of accuracy. So can the rest of us, non-latticists, trust it? On the
one hand, because this result is protected by symmetry the required
precision is not really 3\%. The quantity one must measure is the
deviation from the symmetry limit, $\xi^2-1$, for which the error is
about 25\% and perhaps we should be confident that the lattice result
is correct at this level. On the other hand, this also tells us that
other methods can be competitive at this level. The leading chiral log
calculation\cite{Grinstein:1992qt} gives $\xi\approx1.15$, and the error in
$\xi^2-1$ is estimated from naive dimensional analysis as
$m_K^2/\Lambda_\chi^2\sim24\%$, comparable to the lattice result. 
Moreover, the lattice determination has been made
with only one method (staggered fermions) and it would be reassuring
to see the same result from other methods.  For the lattice to achieve the 0.35\%
accuracy in $\xi$ needed to match the experimental error in
$|V_{td}/V_{ts}|$ a precision of 2\% in the determination of $\xi^2-1$ is required.
Before we, skeptics,  trust any significant improvement in this
determination, other independent
lattice QCD post-dictions of similar accuracy are necessary. 

\subsection{$|V_{cb}|$}
\label{sec:vcb}
\paragraph{Inclusive}
The method of moments gives a very accurate determination of
$|V_{cb}|$ from inclusive semileptonic $B$ decays. In QCD, the rate
${\rm d}\Gamma(B\to X_c\ell\nu)/{\rm d}x\,{\rm d}y=|V_{cb}|^2f(x,y)$, where $x$
and $y$ are the invariant lepton pair mass and energy in units of
$m_B$, is given in terms of four parameters: $|V_{cb}|$, $\alpha_s$, $m_c$
and $m_b$. $|V_{cb}|$, which is what we are after, drops out of
normalized moments. Since $\alpha_s$ is well known, the idea is to fix
$m_c$ and $m_b$ from normalized moments and then use them to compute
the normalization, hence determining $|V_{cb}|$. In reality we cannot
solve QCD to give the moments in terms of $m_c$ and $m_b$, but we can
use a $1/m_Q$ expansion to write the moments in terms of $m_c$, $m_b$
and a few constants that parametrize our
ignorance\cite{Falk:1995kn}. These constants are in fact matrix
elements of operators in the $1/m_Q$ expansion. If terms of order
$1/m_Q^3$ are retained in the expansion one needs to introduce five
such constants; and an additional two are determined by meson
masses. All five constants and two quark masses can be over-determined
from a few normalized moments that are functions of $E_{\rm cut}$, the
lowest limit of the lepton energy integration. The error in the
determination of $|V_{cb}|$ is a remarkably small
2\%\cite{Bauer:2004ve}. But even most remarkable is that this estimate
for the error is truly believable. It is obtained by assigning the
last term {\it retained} in the expansion to the error, as opposed to
the less conservative guessing of the next order {\it not} kept in the
expansion. Since there is also a perturbative expansion, the assigned
error is the combination of the last term kept in all expansions, of
order $\beta_0\alpha_s^2$, $\alpha_s\Lambda_{\rm QCD}/m_b$ and $(\Lambda_{\rm QCD}/m_b)^3$.

There is only one assumption in the calculation that is not fully
justified from first principles. The moment integrals can be computed
perturbatively (in the $1/m_Q$ expansion) only because the integral
can be turned into a contour over a complex $E$ away from the physical
region\cite{Chay:1990da}. However, the contour is pinned at the
minimal energy, $E_{\rm cut}$, on the real axis, right on the physical
cut. So there is a small region of integration where quark-hadron
duality cannot be justified and has to be invoked.  Parametrically
this region of integration is small, a fraction of order $\Lambda/m_Q$ of
the total. But this is a disaster because this is parametrically much
larger than the claimed error of order $(\Lambda/m_Q)^3$. However, this is
believed not to be a problem. For one thing, the fits to moments as
functions of $E_{\rm cut}$ are extremely good: the system is
over-constrained and these internal checks work. And for another, it
has been shown\cite{Boyd:1995ht} that duality works exactly in the
Shifman-Voloshin (small velocity) limit, to order $1/m_Q^2$. It seems
unlikely that the violation to local quark-hadron duality
mainly changes the normalization and has mild dependence on $E_{\rm
cut}$, and that this effect only shows up away from the SV limit.

\paragraph{Exclusive}
The exclusive determination of $|V_{cb}|$ is in pretty good shape
theoretically, but is not competitive with the inclusive one. So it
provides a sanity check, but not an improvement. The 
semileptonic rates into either $D$ or $D^*$ are parametrized by
functions ${\cal F}$, ${\cal F}_*$, of the rapidity of the charmed
meson in the $B$ rest-frame, $w$. Luke's theorem\cite{Luke:1990eg} states 
${\cal F}={\cal F}_*=1+{\cal O}(\Lambda_{\rm QCD}/m_c)^2$ at $w=1$. The rate is
measured at $w>1$ and extrapolated to $w=1$. The extrapolation 
is made with a first principles calculation to avoid introducing
extraneous errors\cite{Boyd:1997kz}. The result has a 4\% error dominated by the
uncertainty in the determination of ${\cal F}$, ${\cal F}_*$ at $w=1$. 

There is some tension between theory and experiment in these exclusive
decays that needs attention. The ratios of form factors $R_{1,2}$ are
at variance from theory by three and two sigma
respectively\cite{Aubert:2006cx}. 
Also, in
the heavy quark limit the slopes $\rho^2$ of ${\cal F}$ and  ${\cal F}_*$
should be equal. One can estimate symmetry violations and obtains\cite{Grinstein:2001yg}
$\rho^2_{{\cal F}}-\rho^2_{{\cal F}_*}\simeq 0.19$, while experimentally this is
$-0.22\pm0.20$, a deviation in the opposite direction. This is a good
place for the lattice to make post-dictions at the few percent error
level that may lend it some credibility in other areas where it
is needed to determine a fundamental parameter.

\subsection{$|V_{ub}|$}
The magnitude $|V_{ub}|$ determines the rate for $B\to X_u\ell\nu$. The well
known experimental difficulty is that since $|V_{ub}|\ll|V_{cb}|$ the
semileptonic decay rate is dominated by charmed final states. To
measure a signal it is necessary to either look at exclusive final
states or  suppress charm kinematically.  The interpretation of the
measurement requires, in the exclusive case, knowledge of hadronic
matrix elements parametrized in terms of form-factors, and for
inclusive decays, understanding of the effect of the kinematic cuts on
the the perturbative expansion and quark-hadron duality.

\paragraph{Inclusive}
This has been the method of choice until recently, since it was
thought that the perturbative calculation was reliable and systematic
and hence could be made sufficiently accurate. However it has become
increasingly clear of late that the calculation cannot be made
arbitrarily precise. The method uses effective field theories to
expand the amplitude systematically in inverse powers of a large
energy, either the heavy mass or the energy of the up-quark (or
equivalently, of the hadronic final state). One shows that in the
restricted kinematic region needed for experiment (to enhance the
up-signal to charm-background) the inclusive amplitude is governed by
a non-perturbative ``shape function,'' which is, however, universal:
it also determines other processes, like the radiative $B\to X_s\gamma$. So
the strategy has been to eliminate this unknown, non-perturbative
function from the rates for semileptonic and radiative decays.

Surprisingly, most analysis do not
eliminate the shape function dependence between the two processes.
 Instead, practitioners
commonly use parametrized fits that unavoidably introduce
uncontrolled errors. It is not surprising that errors quoted in the
determination of $|V_{ub}|$ are smaller if
by a parametrized fit than by the elimination method of
\cite{Leibovich:1999xf}. The problem is that
parameterized fits introduce systematic errors that are unaccounted for. 

Parametrized fits aside, there is an intrinsic problem with the
method. Universality is violated by sub-leading terms\cite{brickwall} in the large
energy expansion (``sub-leading shape functions''). One can estimate
this uncontrolled correction to be of order $\alpha_s\Lambda/m_b$, where $\Lambda$ is
hadronic scale that characterizes the sub-leading effects (in the
effective theory language: matrix elements of higher dimension
operators). We can try to estimate these effects using models of
sub-leading shape functions but then one introduces uncontrolled
errors into the determination. At best one should use models to
estimate the errors. I think it is fair, albeit unpopular, to say that
this method is limited to a precision of about 15\%: since there are
about 10 sub-leading shape functions, I estimate the precision as
$\sqrt{10}\,\alpha_s\Lambda/m_b$. This is much larger than the error commonly
quoted in the determination of $|V_{ub}|$.

This is just as well, since the value of $|V_{ub}|$ from inclusives is
in disagreement not only with the value from exclusives but also with
the global unitarity triangle fit.  You can quantify this if you like,
but it is graphically obvious when you see plots of the fit in the
$\rho$-$\eta$ plane that use only some inputs inputs and contrast
those with the remaining inputs of the global fit. At this conference
last year, Jerome Charles presented\cite{Charles:2006yw} three pairs
of fits contrasting measurements: tree vs.\ loop, CP violating vs.\ CV
conserving, and theory free vs.\ QCD based (see also slide 25 of
Heiko Lacker, this conference). In all these it is evident to the
naked eye that $|V_{ub}|$ (the dark green circle's radius) is too
large; the input used is dominated by inclusives.

\paragraph{Exclusives}
The branching fraction ${\cal B}(B\to\pi\ell \nu) $ is
known\cite{Abe:2004zm} to 8\%. A comparable determination of
$|V_{ub}|$ requires knowledge of the $B\to\pi$ form factor $f_+(q^2)$ to
4\%. There are some things we do know about $f_+$: (i)The shape is
constrained by dispersion relations\cite{Boyd:1994tt}. This means that if we
know $f_+$ at a few well spaced points we can pretty much determine
the whole function $f_+$. (ii)We can get a rough measurement of the
form factor at $q^2=m_\pi^2$ from the rate for $B\to
\pi\pi$\cite{Bauer:2004tj}. This requires a sophisticated effective theory
(SCET) analysis which both shows that the leading order contains a
term with $f_+(m_\pi^2)$ and systematically characterizes the
corrections to the lowest order SCET.  It is safe to assume that
this determination of $f_+(m_\pi^2)$ will not improve beyond the 10\%
mark.

Lattice QCD can determine the form factor, at least over a limited
region of large $q^2$. At the moment there is some disagreement
between the best two lattice calculations, which however use the same
method\cite{Shigemitsu:2004ft}. A skeptic would require not only
agreement between the two existing calculations but also with other
methods, not to mention a set of additional independent successful
post-dictions, before the result can be trusted for a precision
determination of $|V_{ub}|$.

The experimental and lattice measurements can be combined using
constraints from dispersion relations and
unitarity\cite{Arnesen:2005ez}. Because these constraints follow from
fundamentals, they do not introduce additional uncertainties.  They
improve the determination of $|V_{ub}|$ significantly. The lattice
determination is for the $q^2$-region where the rate is smallest. This
is true even if the form factor is largest there, because in that
region the rate is phase space suppressed. But a rough shape of the
spectrum is experimentally observed, through a binned
measurement\cite{Abe:2004zm}, and the dispersion relation constraints
allows one to combine the full experimental spectrum with the
restricted-$q^2$ lattice measurement.  The result of this analysis
gives a 13\% error in $|V_{ub}|$, completely dominated by the lattice
errors.

\paragraph{Alternatives}
Exclusive and inclusive determinations of
$|V_{ub}|$ have comparable precisions. Neither is very good and the
prospect for significant improvement is limited.  Other methods need
be explored, if not to improve on existing $|V_{ub}|$ to lend
confidence to the result. A lattice-free method would be preferable.
A third method, proposed a while ago\cite{Ligeti:1995yz}, uses the
idea of double ratios\cite{Grinstein:1993ys} to reduce hadronic
uncertainties. Two independent approximate symmetries protect double
ratios from deviations from unity, which are therefore of the order of
the product of two small symmetry breaking parameters. For example,
the double ratio
$(f_{B_s}/f_{B_d})/(f_{D_s}/f_{D_d})=(f_{B_s}/f_{D_s})/(f_{B_d}/f_{D_d})=1+{\cal
O}(m_s/m_c)$ because $f_{B_s}/f_{B_d}=f_{D_s}/f_{D_d}=1$ by $SU(3)$
flavor, while $f_{B_s}/f_{D_s}=f_{B_d}/f_{D_d}=\sqrt{m_c/m_b}$ by
heavy flavor symmetry.  One can extract $|V_{ub}/V_{ts}V_{tb}|$ by
measuring the ratio,
\begin{equation}
\frac{{\rm d}\Gamma(\bar B_d\to\rho\ell\nu)/{\rm d}q^2}{{\rm d}\Gamma(\bar B_d\to K^*\ell^+\ell^-)/{\rm d}q^2}
=\frac{|V_{ub}|^2}{|V_{ts}V_{tb}|^2}\cdot\frac{8\pi^2}{\alpha^2}\cdot\frac1{N(q^2)}\cdot
R_B,
\end{equation}
where $q^2$ is the lepton pair invariant mass, and $ N(q^2)$ is a
known function\cite{Grinstein:2004vb}. When expressed as
functions of the rapidity of the vector meson, $y=E_V/m_V$, the ratios
of helicity amplitudes
\begin{equation}
R_B=\frac{\sum_\lambda |H^{B\to\rho}_\lambda(y)|^2}{\sum_\lambda
  |H^{B\to K^*}_\lambda(y)|^2},\quad
R_D=\frac{\sum_\lambda |H^{D\to\rho}_\lambda(y)|^2}{\sum_\lambda |H^{D\to K^*}_\lambda(y)|^2},\nn
\end{equation}
are related by a double ratio: $R_B(y)=R_D(y)(1+{\cal
  O}(m_s(m_c^{-1}-m_b^{-1})))$. 
This measurement could be done today:  CLEO has accurately measured the
  required semileptonic $D$ decays\cite{Adam:2007pv,Gray:2007pw}.

A fourth method is available if we are willing to use rarer
decays. To extract $|V_{ub}|$ from  ${\cal B}(B^+\to\tau^+\nu_\tau)
=(0.88{\textstyle{+0.68\atop-0.67}}\pm0.11)\times10^{-4}$\cite{Aubert:2004kz} one
needs a lattice determination of $f_B$. Since we want to move away
from relying on non-perturbative methods (lattice) to extract $|V_{ub}|$
we propose a cleaner but more difficult measurement, the double ratio
\begin{equation}
\frac{\frac{\Gamma(B_u\to\tau\nu)}{\Gamma(B_s\to\ell^+\ell^-)}}{\frac{\Gamma(D_d\to\ell
    \nu)}{\Gamma(D_s\to\ell\nu)}}\sim
\frac{|V_{ub}|^2}{|V_{ts}V_{tb}|^2}\cdot\frac{\pi^2}{\alpha^2}\cdot\left(\frac{f_B/f_{B_s}}{f_D/f_{D_s}}\right)^2
\end{equation}
In the SM ${\cal B}(B_s\to\mu^+\mu^-)\approx 3.5\times10^{-9}$ $\times
 (f_{B_s}/210\,\text{MeV})^2(|V_{ts}|/0.040)^2$
is the only presently unknown quantity in the double ratio and is
expected to be well measured at the LHC\cite{Schopper:2006he}. 

The ratio $\Gamma(B^+\to\tau^+\nu)/ \Gamma(B_d\to\mu^+\mu^-)$ gives us a fifth method. It
has basically no hadronic uncertainty, since the hadronic factor
$f_B/f_{B_d}=1$, by isospin. It involves$|V_{ub}|^2/|V_{td}V_{tb}|^2$,
an unusual combination of CKMs. In the $\rho-\eta$ plane it forms a circle
centered at $\sim (-0.2,0)$ of radius $\sim0.5$. Of course, measuring
$\Gamma(B_d\to\mu^+\mu^-)$ is extremely hard. 

In a sixth method one
studies wrong charm decays $\bar B_{d,s}\to\bar DX$ (really $b\bar q\to u
\bar c$). This can be done both in semi-inclusive
decays\cite{Falk:1999sa} (an experimentally challenging
measurement) or in exclusive decays\cite{Evans:1999wx} (where
an interesting connection to $B_{d,s}$ mixing matrix elements is
involved).

\subsection{$\alpha$ from $B\to\pi\pi,\pi\rho,\rho\rho $.}  In
principle the penguin contamination problem\cite{Grinstein:1989df}
requires a full isospin analysis\cite{Gronau:1990ka} for a
theoretically clean determination of the angle $\alpha$.  The angle
determination works slightly better than we had a right to expect a
priori.  The reason lies in two empirical observation in $B\to\rho\rho$.
First, the longitudinal polarization dominates, and therefore the
final state is to good approximation a CP eigenstate (CP even, in
fact). And second, the branching fraction for $B\to\rho^0\rho^0 $ is small:
relative to $B\to\rho^+\rho^- $ it is $6\pm3\%$, to be compared with the neutral
to charged decay into pions of $23\pm4\%$. This means that the
contamination from penguin operators is small and one can get a clean
measurement of $\alpha$. All three decay modes are about equally important
in current fits, which give $\alpha=93{\textstyle{+11\atop-9}}$ degrees.

\subsection{$\gamma$ from $B^\pm\to DK^\pm $.}
Three different methods are used. They are all based on the
interference between Cabibbo-allowed ({\it e.g.,} $B^-\to D^0K^-$) and
suppressed decays ({\it e.g.,} $B^-\to \overline{D^0}K^-$) with $ D^0$,
$\overline{D^0}$ decaying to a common state. The
GLW\cite{Gronau:1990ra} method uses decays to a common CP
eigenstate. In the ADS method\cite{Atwood:1996ci} the final state is
chosen to be a suppressed $D$ decay mode if the $D$ came from an
allowed $B$ decay; for example, the final state in the charm decay can
be taken to be $K^+\pi^-$ so it is doubly Cabibbo suppressed for a $D^0$
decay but allowed for a $\overline{D^0}$ decay. The efficacy of this
method depends sensitively on the ratio of amplitudes,
which can be measured separately, $r_B=|A(B^-\to
\overline{D^0}K^-)/A(B^-\to D^0K^-)|$. In the GGSZ
method\cite{Giri:2003ty} the $D^0$ and $\overline{D^0}$ are
reconstructed in a common three body final state. The results to date vary depending
on which decay mode is actually used, so the determination of $\gamma$ from
all measurements combined is not very good,
$\gamma=62{\textstyle{+38\atop-24}}$ degrees. More data should improve the
determination of $\gamma$.

\subsection{Are there anomalies?}
There seem to be as many papers in the literature claiming there is a
``$B\to K\pi$ puzzle'' as those that claim it is not a puzzle. It is
easy to see why. In order to find a puzzle one must know a priori the
hadronic amplitudes. Those who find a puzzle in $B\to K\pi$ make
assumptions about hadronic amplitudes that those who find no puzzle
think are unwarranted.  Moreover, Ref.~\cite{donoghue} showed that
soft final state interactions  do not disappear in the large $m_b$
limit, and Refs.~\cite{falkFSI} and~\cite{wolfensteinFSI} studied this
quantitatively for $B\to K\pi$ and $B\to\pi\pi $, respectively, and concluded
the effects should be expected to be large. For example, the CP
asymmetry in $B\to K\pi$ could easily be 20\% and the bound $\sin^2\gamma\leq R$,
where $R=\Gamma(B_d\to\pi^\mp K^\pm )/ \Gamma(B^\pm\to\pi^\pm K)$ could easily be violated at
the 20\% level. 

The case for new physics in CPV in charmless $b\to s$ decays would seem
to be stronger. Regardless of decay mode  $\beta_{eff}$ is predicted by
SCET, QCD-factorization and pQCD to deviate from $\beta_{J/\psi K_s}$ by a small
positive amount. Experimentally the deviations vary from mode to mode
but are all non-positive  and not necessarily small. However, many things
have to be checked before one can begin to believe we are seeing new
physics here. First, all of the theoretical schemes need to come to
terms with the soft final state interactions  issue raised in
\cite{donoghue} or show that work is incorrect. Then, also, the fact
that all deviations are negative strongly suggests that the
measurements have been corrupted by an admixture of the opposite CP
final state. 

In my view there is at present no case for deviations from the standard CKM
model of flavor.

\section{Perspective}

How precise should we ultimately measure the elements of the
CKM matrix? I am not asking what is the ultimate
precision afforded by present day methods, but rather, how precisely
do we need to know them. A rather common answer is that one should
aspire to determine them as well as possible given available methods
because the CKM elements are fundamental constants of nature, as
fundamental as any other coupling in the Lagrangian of the Standard
Model of electroweak and strong interactions (SM). But I find this
answer lame and na\"\i{}ve, particularly when the effort is rather
expensive both in real money and in human capital. A much better
answer is obtained by estimating realistically how large deviation due
to new physics could reasonably be. 

It is not difficult to
find extensions of the standard model that would give deviations from
expected measurements just beyond the precision attained to date. For
example, one can take the minimal extension to the supersymmetrized SM
(the MSSM), and choose parameters appropriately, that is, on the verge
of being ruled out (or discovered). But this is contrived, and 
not  a reasonable way to answer our question. 

One way of estimating the precision with which we need to determine
CKM elements is to verify that the CKM matrix is unitary. Violations
to CKM unitarity must come from additional quarks beyond those in the
SM. This is already very constrained by electroweak precision measurements and
for that reason I will not consider it any further (but creative theorists can
get around these constraints; see, {\it e.g.,} Ref.~ \cite{Hou:2005yb}).

Instead I will concentrate on the question, which I think is more
interesting, what precision is needed to exclude new physics at the
TeV scale? 
%
In the absence of new dynamics radiative
corrections would render the mass scale of the electroweak theory
comparable to the Planck  scale. New physics at the TeV scale is
generally invoked to explain this ``hierarchy problem.'' But quark
mass terms break the electroweak symmetry group, so the quark mass
matrices are necessarily connected to this new physics. New
``higgs dynamics'' at the TeV scale must incorporate new flavor
physics too.

This suggests another criterion for the required precision in the
determination of CKMs, namely, enough that we can see clearly the
effects of this new flavor physics originating from the new, TeV-scale
dynamics. It is easy to describe the effects of new TeV dynamics at
below TeV energies in a model independent way. One simply extends the
Lagrangian of the SM by operators of dimension higher than four,
suppressed by powers of the new physics scale, $\Lambda$. The work in
\cite{Buchmuller:1985jz,Leung:1984ni} lists all operators of dimension
five and six and analyzes some of their effects. Ignoring operators
mediating flavor changing neutral currents (FCNC), $\Lambda\sim$ a few TeV is
consistent with experiment. But if the coefficient of FCNC operators
is given by dimensional analysis, then $\Lambda\sim$ a few TeV is strongly
excluded. A much larger scale, $\Lambda\sim 10^4$~TeV, is still consistent with
experiment, but then a hierarchy problem reappears.

Let $\CA$ denote the amplitude for some process which we
write as the sum of SM and new physics pieces,
$\CA=\CA_{\text{SM}}+\CA_{\text{New}}$. If this proceeds at tree level
in the SM we estimate, roughly, 
\begin{equation}
\CA_{\text{SM}}\sim \frac{g^2}{M_W^2}\times\text{CKM}\quad\text{and}\quad
\CA_{\text{New}}\sim \frac{1}{\Lambda^2},
\end{equation}
where the factor ``CKM'' stands for some combination of CKM
elements. If we want to be sensitive to the the second term the
uncertainty in the first one should be no larger than the expected
size of new physics effects:
\begin{equation}
\label{eq:treePC}
\frac{\delta(\text{CKM})}{\text{CKM}} \sim
\frac1{\text{CKM}}\frac{1/\Lambda^2}{g^2/M_W^2}
\sim1\%\!\left(\frac{0.03}{\text{CKM}}\right)
     \!\!    \left(\frac{10~\text{TeV}}{\Lambda}\right)^{\!\!2}
\end{equation}

Repeat now the power counting leading to \eqref{eq:treePC}, but for processes
involving FCNC. These require at least one loop in the SM, but not in
the new physics. We now estimate
\begin{equation}
\CA_{\text{SM}}\sim \frac{\alpha}{4\pi\sin^2\theta_w}\frac{g^2}{M_W^2}\times\text{CKM},
\end{equation}
so that 
\begin{align}
\label{eq:loopPC}
\frac{\delta(\text{CKM})}{\text{CKM}} &\sim
\frac1{\text{CKM}}\frac{1/\Lambda^2}{({\alpha}/{4\pi\sin^2\theta_w})(g^2/M_W^2)}\nn\\
&\sim400\% \times\left(\frac{0.03}{\text{CKM}}\right)
           \left(\frac{10~\text{TeV}}{\Lambda}\right)^{\!\!2}
\end{align}
This is an underestimate since for SM's FCNC the CKM combination is
smaller than 0.03. Alternatively one can write this as a limit one
places on the scale of new physics,
\eqref{eq:loopPC} gives
\begin{multline}
\label{eq:loopPCsolved}
\Lambda > v~\sqrt{\frac{1}{\frac{\delta(\text{CKM})}{\text{CKM}}}\frac1{\text{CKM}}\frac{4\pi\sin^2\theta_w}{\alpha}
}\\
\sim10^3~\text{TeV}\times\left(\frac{10\%}{\frac{\delta(\text{CKM})}{\text{CKM}}}\right)^{\frac12}
\left(\frac{0.0002}{\text{CKM} }\right)^{\frac12}
\end{multline}
So 10\% precision already makes a strong statement about the scale of
new physics, $\Lambda$. We argued above that since the solution to the
hierarchy problem involves the higgs (or more generally, the breaking
of EW symmetry), and since this is responsible for quark/lepton
masses, then it is natural that the new physics that solves the
hierarchy involves flavor. 

What gives? I see  three possibilities:
\begin{enumerate}
\item Cancellations among several new physics (NP) contributions
\item Large scale ($\Lambda\sim1000$~TeV) of NP except for a light higgs
  (or a light flavor blind sector)
\item Automatic alignment of small CKMs in NP with the small CKMs
  in SM. 
\end{enumerate}

Let's examine these generic possibilities a bit more closely. The
first one presumes there is new physics at the TeV scale. It explains
the absence of FCNCs by happenstance, masses and couplings in the sum
of terms contributing to the amplitude conspiring to cancel to good
approximation. When stated this way, this seems like a very
unappealing possibility. Yet this is what the MSSM does, particularly
if one insists on light ($\sim100$~ GeV) partners of ``normal''
particles. For example, in order to avoid unacceptably large rate for
$B\to X_s\gamma$, it needs to cancel the charged higgs mediated
contribution (which always adds coherently to the SM contribution and
is only suppressed to the extent that the charged higgs mass is made
heavy) against some other contribution, like a gaugino mediated
graph. In general terms, if we are willing to allow some level of fine
tuning to suppress FCNC then we correspondingly loose a clear idea of
what is the scale we are probing, or equivalently, what we are aiming
at.

The second possibility is that the scale of the flavor dynamics is
about 1000 TeV (or larger). FCNCs associated with this
scale are not experimentally ruled out. Yet the scale of EW symmetry
breaking is three orders of magnitude smaller, so this is a new, but
smaller, hierarchy problem. One expects dynamics that solves the
hierarchy problem to show up at the LHC (either the little hierarchy
or the big one or both), but, depending on the actual scale of flavor,
there may be no sign of FCNCs in $B$ and $K$ physics. Technicolor
models in which flavor is generated by an extended sector at the
1000~TeV scale fall in this class, as do  many more modern  examples
of theories designed to solve the little hierarchy problem; see, for
example, Refs.~\cite{Arkani-Hamed:2002qx,Cheng:2004yc,Chacko:2005pe}.

The third and last possibility is that the NP at the TeV scale is
aligned in flavor with the SM. The reason FCNCs are suppressed in the
SM is that they do not appear at tree level and they are suppressed by
a small CKM factor. The NP is not ruled out if it has the same (or
similarly suppressed) CKM factors associated with FCNCs. The difference
with the first possibility is that no cancellation of graphs is
required, other than those cancellations that follow automatically
from the unitarity of the CKM matrix. Indeed, we see from
Eq.~\eqref{eq:loopPCsolved} that if we take away the last factor the
scale of new physics is only bounded to be greater than about
10~TeV. In fact SUSY theories make use of this automatic suppression,
and are free of additional fine tunings if one can take all SUSY
masses to be $\sim10~{\rm TeV}/4\pi\sim1~{\rm TeV}$. The first possibility
discussed above applies to SUSY if one insists that SUSY masses are
much lighter, say, with masses of a few hundred GeV. This third and
last possibility is appealing in the sense that it makes fairly
definite predictions and should be accessible experimentally.

This is made even more appealing by realizing that it follows
naturally from imposing a simple principle based on symmetry
considerations alone. In the absence of quark masses the SM has a
large flavor symmetry, $SU(3)^3$ (one factor of $SU(3)$ for each of
quark doublets, right handed up-type quarks and right handed down-type
quarks).  The principle of Minimal Flavor Violation asserts that this
symmetry is only violated by the quark mass matrices. Any new
interaction that breaks this large flavor symmetry must do so by
including the appropriately transforming combination of quark mass
matrices. This can be implemented as an effective theory, by adding
higher dimension operators to the SM suppressed by powers of the NP
scale $\Lambda$, as discussed above. The difference is that now the
coefficients of these operators are the product of an unknown constant
of order one times a factor of the quark mass matrix fixed by these
symmetry considerations. In the quark mass eigenstate basis this gives
rise to coefficients that include small CKM suppression factors in
FCNCs. A complete analysis of the effects of dimensions six operators
on FCNCs has been performed\cite{MFV} and shows that the scale of NP must be of
the order of 10~TeV, in accordance with the crude estimates
above. The most stringent bound comes from radiative $B$
decays ($\Lambda\geq 9$~TeV), with other processes giving bounds in the range
1~TeV to 6~TeV. I believe the aim of FPCP should be to exclude
$\Lambda\leq10$~TeV in MFV from all FCNC processes.

There exist other mechanisms, like next-to-minimal Flavor Violation,
which also naturally produce small coefficients for NP contributions to
FCNCs. Since MFV gives the minimal expected deviations of FCNC from SM
predictions it still serves as a template against which one should
calibrate experimental reach. For more on these alternatives see Ref.~\cite{NMFV}.

If $\Lambda<10$~TeV MFV is excluded then one should expect that $\Lambda>10$~TeV
also for flavor conserving NP. If NP is found at the LHC (say, as
anomalous higgs or $W$ couplings), it would be strongly suggestive
that the scale of FP is large, $\Lambda_{FP}>1000$~TeV. Although this would
be bad news for this workshop, it would be very interesting as it
would suggest that the second possibility above is the correct
one. The LHC would then explore the physics of EW symmetry breaking
(higgs properties, perhaps techniparticles) and we would have to be
creative to figure out how to explore the much higher scale of flavor
physics. 

Alternatively, if deviations from SM FCNCs are found and are
consistent with MFV (or its extensions) with $\Lambda\sim10$~TeV then for
weakly coupled NP the new particles have masses of the order of a few
TeV. This could be just beyond the reach of the LHC. I can't help but
pointing out that this would have been well within the reach of the
SSC! In any case, FPCP would afford the best look at physics beyond
the SM. 

MFV has many surprising implications. But none is more striking than
the following. If  leptons and quarks unify, and if the
solution to the hierarchy problem introduces flavor physics at the TeV
scale then\cite{MFVGUTS} lepton flavor  violation should be observed in $\mu\to e$
processes at MEG and PRISM. Exciting flavor physics ahead, indeed!

\begin{acknowledgments}
Work supported in part by the Department of Energy under contract DE-FG03-97E
R40546. 
\end{acknowledgments}

\bigskip 

\end{document}